\documentclass[twocolumn]
{aastex631}
\usepackage{multirow}
\usepackage{times}

\newcommand{\xmm}{\textit{XMM-Newton}}

\makeatletter
\let\old@cite\cite 
\renewcommand{\cite}[1]{%
  \begingroup
  \let\NAT@open\@empty 
  \let\NAT@close\@empty 
  \old@cite{#1}%
  \endgroup
}
\makeatother
\usepackage{amsmath}
\graphicspath{{./}{figures/}}
\shortauthors{WEN et al.}

\begin{document}

\title{Mrk~382: A Narrow-line Seyfert 1 Galaxy with Recurrent X-ray State Transitions}

\author[0000-0003-4897-4106]{YanLi Ai}
\affiliation{Shenzhen Key Laboratory of Ultraintense Laser and Advanced Material Technology, Center for Intense Laser Application Technology, and College of Engineering Physics, Shenzhen Technology University, Shenzhen 518118, China}

\correspondingauthor{YanLi Ai}
\email{aiyanli@sztu.edu.cn}

\author[0009-0003-5419-7355]{WenFeng Wen}
\affiliation{Shenzhen Key Laboratory of Ultraintense Laser and Advanced Material Technology, Center for Intense Laser Application Technology, and College of Engineering Physics, Shenzhen Technology University, Shenzhen 518118, China}

\author[0000-0002-4757-8622]{Liming Dou}

\affiliation{Department of Astronomy, Guangzhou University, Guangzhou 510006, China} 

\author[0009-0003-1518-6186]{Jiahua Wu}
\affiliation{Department of Astronomy, Guangzhou University, Guangzhou 510006, China} 

\author{Chen Hu}
\affiliation{Key Laboratory for Particle Astrophysics, Institute of High Energy Physics, Chinese Academy of Sciences,
 19B Yuquan Road, Beijing 100049, China}

\author[0000-0002-1517-6792]{Tinggui Wang}
\affiliation{School of Astronomy and Space Sciences, University of Science and Technology of China, Hefei, People's Republic of China}
\affiliation{Key Laboratory for Research in Galaxies and Cosmology of Chinese Academy of Sciences, Department of Astronomy, University of Science and Technology of China, Hefei, China}

\author[0009-0007-7426-2758]{XiaoHui Yang}
\affiliation{Shenzhen Key Laboratory of Ultraintense Laser and Advanced Material Technology, Center for Intense Laser Application Technology, and College of Engineering Physics, Shenzhen Technology University, Shenzhen 518118, China}

\author{Jing Wang}
\affiliation{National Astronomical Observatories, Beijing 100101, China}

\author[0000-0002-7350-6913]{Xue-Bing Wu} 
\affiliation{Department of Astronomy, School of Physics, Peking University, Beijing 100871, China}
\affiliation{Kavli Institute for Astronomy and Astrophysics, Peking University, Beijing 100871, China}

\author{Qiusheng Gu}
\affiliation{School of Astronomy and Space Science, Nanjing University, Nanjing 210093, China}
\affiliation{Key Laboratory of Modern Astronomy and Astrophysics (Nanjing University), Ministry of Education, Nanjing 210093, China}

\author{Xinwen Shu}
\affiliation{ Department of Physics, Anhui Normal University, Wuhu, Anhui, 241000, China}

\author[0000-0002-5830-3544]{PuDu}
\affiliation{Key Laboratory for Particle Astrophysics, Institute of High Energy Physics, Chinese Academy of Sciences,
 19B Yuquan Road, Beijing 100049, China}

\author[0000-0001-9449-9268]{Jian-Min Wang}
\affiliation{Key Laboratory for Particle Astrophysics, Institute of High Energy Physics, Chinese Academy of Sciences,
 19B Yuquan Road, Beijing 100049, China}

\begin{abstract}
We report recurrent X-ray state transitions in the nearby narrow-line Seyfert 1 galaxy Mrk 382 using multi-epoch observations from Swift, Chandra, XMM-Newton, and eROSITA, together with archival ultraviolet, optical, and infrared data. The 0.3–2 keV flux varies by nearly an order of magnitude over $\sim15$ yr, with multiple transitions between bright and faint states. The X-ray spectrum shows strong state-dependent evolution, changing from a steep high-state continuum ($\Gamma=2.32\pm0.04$) to a much harder low-state spectrum ($\Gamma=1.39\pm0.06$), which also exhibits a narrow Fe K$\alpha$ line with an equivalent width of $\sim330$ eV. Reflection modeling indicates that the low-flux state is strongly reflection dominated, with the reflection fraction increasing from $R_{\rm refl}\sim4.4$ to $\sim30$, consistent with substantial changes in the inner disk–corona configuration. Photoionized absorption contributes to the spectral complexity, but its relatively modest column density suggests that absorption alone is unlikely to explain the full amplitude of the X-ray transitions. The ultraviolet emission shows possible long-term correspondence with the X-ray variability at smaller amplitude, while the optical and mid-infrared bands vary more mildly. The spectral and multiwavelength behavior suggests that changes in coronal geometry may be accompanied by changes in its intrinsic energetics. Despite the dramatic X-ray variability, no extreme X-ray-weak state or clear optical spectral-type change is observed. Mrk 382 therefore provides a valuable nearby laboratory for studying the evolution of the disk–corona system and its connection to multiwavelength AGN variability.

\end{abstract}


\section{Introduction} \label{sec:intro}

The X-ray emission of radio-quiet active galactic nuclei
(AGNs) is generally understood to originate from inverse
Compton scattering of optical/UV photons in a hot corona
above the accretion disk
(e.g., \citealt{sunyaev1980,yuan2014}).
X-ray variability therefore provides a powerful probe of the
physical conditions in the corona and the innermost accretion
flow. In most AGNs, the observed X-ray flux variability is
relatively modest, typically within a factor of $\sim$2
(e.g., \citealt{gibson2012,middei2017,timlin2020}).
By contrast, extreme X-ray variability with amplitudes exceeding
a factor of $\gtrsim$10 remains uncommon in the general AGN
population (e.g., \citealt{bi2015,medvedev2022,wangH2025}).

Although extreme X-ray variability is relatively rare among AGNs,
a growing number of such sources have been identified in recent
years, enabling detailed investigations of their physical origins.
Particularly relevant in this context are changing-look (CL) AGNs,
which exhibit dramatic transitions in both their X-ray emission
and optical spectral type, often interpreted as signatures of
substantial changes in the inner accretion flow
(e.g., \citealt{grupe2015,krumpe2017,ai2020,liu2020,jana2021,yang2023}).
Among them,  a small number of CL AGNs have been found to exhibit recurrent
CL behavior, inferred from sparse spectroscopic observations
spanning several decades, with transition timescales ranging from
months to years \citep{komossa2026}.
In these systems, large-amplitude X-ray variability is frequently
accompanied by pronounced optical spectral evolution, suggesting that
the structure of the disk--corona system can undergo substantial
reconfiguration on relatively short timescales, in some cases even
within months.

However, not all AGNs with extreme X-ray variability display
classical CL behavior. In some type~1 AGNs, large-amplitude
X-ray variability has been associated with changes in the column
density of dust-free obscuring material along the line of sight and covering the central AGN
(e.g., \citealt{wangyj2022,mehdipour2021, Huang2026}),
while in other cases the variability appears to be intrinsic to
the X-ray emitting corona, likely driven by changes in its geometry
or energetics (e.g., \citealt{Miniutti2012,Ricci2020,Wu2020}).

Strong and sometimes rapid X-ray variability has also been
observed in narrow-line Seyfert~1 galaxies (NLS1s), which are
generally characterized by high or even super-Eddington
accretion rates
(e.g., \citealt{ai2011,reeves2019,boller2021,
parker2021,jin2023}).
Their high accretion rates
are expected to produce unstable disk--corona configurations and
powerful radiatively driven winds, making them especially prone
to strong and complex X-ray variability (e.g.,  \citealt{jiang2019global};
\citealt{yang2025wind}). Recent studies have shown that such
CL behavior can also occur in high-accretion
NLS1 galaxies, suggesting that NLS1s may represent an
important subclass of highly variable AGNs operating close
to the Eddington limit
\citep{macleod2019, frederick2019ZTF18aajupnt, hon2022J1406,parker2019bgc1566,xu2024}.

Repeated transitions between X-ray bright and faint states
have been observed in several highly variable AGNs and CL
systems. For example, multiple CL events have
been reported in Mrk~590 \citep{denny2014mrk509,
palit2026}. Among nearby NLS1 galaxies,
NGC~1566 represents one of the nearest known repeating CL
AGNs, exhibiting dramatic multiwavelength variability
associated with recurrent state transitions \citep{parker2019bgc1566,xu2024}. However, there
also exist sources that show strong recurrent X-ray variability
without clear optical spectral-type transitions, such as
Mrk~335. These objects are sometimes referred to as
``frozen-look'' AGNs, and their physical nature remains
uncertain \citep{komossa2026}. Although CL behavior has also been identified in
several NLS1 galaxies, recent studies suggest that classical
CL transitions preferentially occur in systems with relatively
low Eddington ratios, typically
$\lambda_{\rm Edd}\lesssim0.01$
\citep{wangs2024}. In addition, the critical accretion
threshold required for the CL phenomenon may
depend on black hole mass, with lower-mass supermassive
black holes requiring higher Eddington ratios to fully
suppress the broad-line region \citep{Guo2025}. These
results imply that the physical origin of extreme
variability in highly accreting NLS1 galaxies may differ
from that in typical CL AGNs.

In this work, we present a detailed analysis of the NLS1 galaxy
Mrk~382, which exhibits recurrent and extreme X-ray variability.
As a nearby narrow-line Seyfert~1 galaxy at $z=0.0332$ with a
high Eddington ratio, Mrk~382 provides an excellent laboratory
for investigating the physical origin of recurrent X-ray state
transitions in highly accreting AGNs. Mrk~382 is also one of the
targets in the reverberation-mapping (RM) campaign of
super-Eddington accreting massive black holes (SEAMBHs),
with a black hole mass of
$\log(M_{\rm BH}/M_\odot)=6.50^{+0.19}_{-0.29}$ and 
a super-Eddington accretion rate
($\dot{m}\sim6$, e.g., \citealt{du2015,hu2015,Tortosa2023}). Previous X-ray studies based on \textit{XMM-Newton} and \textit{NuSTAR} observations revealed strong relativistic
reflection features together with pronounced spectral
variability in Mrk~382, with the source transitioning from a
reflection-dominated low state to a continuum-dominated high
state \citep{xu2025}. These results suggest that
the observed variability is closely related to changes in the
geometry and compactness of the inner corona, likely regulated
by strong gravitational light-bending effects near the black
hole.

Here, we present a comprehensive multiwavelength investigation of
Mrk~382, combining long-term X-ray monitoring from
\textit{Swift}, \textit{Chandra}, \textit{XMM-Newton}, and
eROSITA with ultraviolet, optical, and infrared observations.
By jointly analyzing the spectral and temporal behavior across
different wavebands, we aim to constrain the physical origin of
its recurrent X-ray variability. This paper is organized as follows. In Section~\ref{sec:2},
we describe the X-ray and multiwavelength observations.
In Section~\ref{sec:X-RAY AND THE MUTILWAVELENGTH PROPERTIE},
we present the spectral and variability analysis results.
In Section~\ref{sec:D}, we discuss the physical implications.
Finally, in Section~\ref{sec:C}, we summarize our conclusions.
Throughout this paper, we adopt a standard $\Lambda$CDM
cosmology with $\Omega_\Lambda = 0.7$, $\Omega_m = 0.3$,
and $H_0 = 70$ km s$^{-1}$ Mpc$^{-1}$.

\section{Observations and Data Reduction} \label{sec:2}
Mrk 382 has archival \textit{Swift} and {\xmm} observations, which we complement with new \textit{Swift} XRT/UVOT monitoring from our dedicated program (Sw8--Sw13; Table~\ref{tab:observation}). \textit{Chandra} and eROSITA observations are also included.

\subsection{XMM-Newton Observations}

Two {\xmm} observations were performed on 2011 November 02 and 2019 October 30, hereafter referred to as XM\_H and XM\_L, respectively.
This study utilizes EPIC-PN data for spectral analysis due to its superior signal-to-noise ratio compared to the EPIC-MOS data. For the 2011 observation, the PN camera was operated in full window mode with
a thin filter, while for the 2019 observation, it was in small window
mode with the same filter. 

The EPIC-pn data were processed with the {\xmm} Science Analysis System (SAS v21.0.0; \citealt{gabriel2004}) and the latest calibration files, following standard procedures for point sources. Only single and double events (PATTERN $\leq$ 4) were selected, and periods of high background were removed. We checked for pile-up using the SAS task \textit{epatplot}. Pile-up was detected in the XM\_H spectrum. To exclude the piled-up core, the XM\_H source spectrum was extracted from an annular region with inner and outer radii of $9^{\prime\prime}$ and $30^{\prime\prime}$, respectively. The XM\_L spectrum was extracted from a circular region with a radius of $30^{\prime\prime}$. Background spectra were extracted from source-free regions on the same CCD chip with a radius of $60^{\prime\prime}$. All spectra were grouped using \textit{grppha} with a minimum of 100 counts per energy bin.

The Optical Monitor (OM) observations were obtained with the UVW2, UVM2, and UVW1 filters, with effective wavelengths of 2120~\AA, 2310~\AA, and 2910~\AA, respectively. The OM data were reduced using the task \textit{omichain}. Photometric measurements from individual exposures were extracted from the SWSRLI files, and mean magnitudes and fluxes were adopted for each filter. 


\subsection{Swift Observations}
Mrk 382 was observed by the \textit{Neil Gehrels Swift Observatory}
\citep{gehrels2004} 13 times between 2009 and 2025 (see
Table~\ref{tab:observation}). The earlier observations (Sw1--Sw7) were
obtained from the archival \textit{Swift} data, while the later
observations (Sw8--Sw13) were carried out as part of our dedicated
\textit{Swift} XRT/UVOT monitoring program. XRT data are available for
all epochs, whereas simultaneous UVOT observations exist only for a
subset of them.

The XRT data were analyzed in photon-counting (PC) mode. Spectra were
extracted using \textit{xselect} (version 2.5b). All observations have
count rates below 0.4 counts s$^{-1}$, well within the pile-up--free
threshold of 0.5 counts s$^{-1}$ for PC mode. Source photons were
extracted from a circular region centered on the optical position of
Mrk 382 with a radius of $47^{\prime\prime}$, while background photons
were extracted from a nearby source-free circular region with a radius
of $94^{\prime\prime}$. Ancillary response files were generated using
\textit{xrtmkarf}, and the redistribution matrix files were obtained
from CALDB. The spectra were grouped using \textit{grppha} to contain
at least two counts per bin.

For the epochs with simultaneous UVOT observations, data were obtained
using filters similar to those of the {\xmm} OM (UVW1, UVW2, and UVM2;
see Table~\ref{tab:observation}). The task \textit{uvotdetect} was used
to determine the source position. Source counts were extracted from a
circular region with a radius of $5^{\prime\prime}$ centered on the
source, and background counts were extracted from a nearby source-free
circular region with a radius of $20^{\prime\prime}$. Magnitudes and
flux densities were computed using \textit{uvotsource}.

\subsection{eROSITA observations}

eROSITA aboard the Spectrum--RG (SRG) observatory \citep{sunyaev2021}
is performing an all-sky X-ray survey. Mrk 382 was observed once on
2020 April 18. The observation details are listed in
Table~\ref{tab:observation}. We used the standard pipeline (version 010)
data products, combining all seven telescope modules (TM1--TM7).

\subsection{Chandra Observations}
Mrk 382 was observed with \textit{Chandra} using ACIS-S
\citep{garmire2003} on 2010 December 18 (see
Table~\ref{tab:observation}). The Chandra observation analyzed in this work is available through
\dataset[DOI: 10.25574/cdc.625]{https://doi.org/10.25574/cdc.625}. The data were processed with the
\textit{Chandra} Interactive Analysis of Observations (CIAO; v4.16;
\citealt{fruscione2006}) using CALDB 4.11.3. We first ran
\texttt{chandra\_repro} to generate a new level-2 event file.
Background flares were filtered using \texttt{deflare} with an
iterative $3\sigma$ clipping algorithm, resulting in a cleaned exposure
time of 4.6 ks.

The source spectrum was extracted using \texttt{specextract} from a
circular region with a radius of $3^{\prime\prime}$ centered on the
X-ray position. The background spectrum was extracted from an annular
region centered on the source with inner and outer radii of
$5^{\prime\prime}$ and $10^{\prime\prime}$, respectively. 
The spectrum was grouped to contain at least 20 counts per bin for
spectral fitting.

\begin{deluxetable*}{cccccc}
\tabletypesize{\scriptsize}
\tablecaption{X-ray and UV observation log of Mrk 382.}
\tablenum{1}
\label{tab:observation}
\tablewidth{0pt}
\tablehead{
\colhead{Observatory} &
\colhead{ObsID} &
\colhead{Date} &
\colhead{Exposure (ks)$^{a}$} &
\colhead{Abbr.} &
\colhead{UV filters}
}
\startdata
\textit{Swift} XRT & 00038126001 & 2009-02-26 & 3.1 & Sw1 & \\
\textit{Swift} XRT & 00038126002 & 2009-08-31 & 2.1 & Sw2 & UVW2 \\
Chandra ACIS & 13008 & 2010-12-06 & 4.6 & Ch1 & \\
\textit{Swift} XRT & 00040527001 & 2011-08-31 & 0.1 & Sw3 & \\
\textit{Swift} XRT & 00040527002 & 2011-09-01 & 4.6 & Sw4 & UVW2, UVM2, UVW1 \\
\textit{XMM-Newton} & 0670040101 & 2011-11-02 & 25.8 & XM\_H & UVW1, UVM2 \\
\textit{Swift} XRT & 00088828001 & 2019-10-29 & 1.5 & Sw5 & UVW2 \\
\textit{XMM-Newton} & 0843020801 & 2019-10-30 & 35.7 & XM\_L & UVM2, UVW1 \\
eROSITA (TM1--7) & -- & 2020-04-18 & 0.2 & eR1 & \\
\textit{Swift} XRT & 00088828003 & 2021-01-01 & 2.4 & Sw6 & UVW2, UVM2, UVW1 \\
\textit{Swift} XRT & 00088828004 & 2021-01-03 & 3.1 & Sw7 & UVW2, UVM2, UVW1 \\
\textit{Swift} XRT & 00019723001 & 2025-04-24 & 1.7 & Sw8 & \\
\textit{Swift} XRT & 00019723002 & 2025-05-22 & 0.9 & Sw9 & UVW2 \\
\textit{Swift} XRT & 00019723005 & 2025-09-13 & 1.4 & Sw10 & UVW1 \\
\textit{Swift} XRT & 00019723006 & 2025-09-29 & 0.7 & Sw11 & \\
\textit{Swift} XRT & 00019723008 & 2025-10-11 & 0.7 & Sw12 & \\
\textit{Swift} XRT & 00019723009 & 2025-12-11 & 0.7 & Sw13 & \\
\enddata
\tablenotetext{a}{Exposure denotes the net good-time interval after screening.}
\end{deluxetable*}

\begin{figure}
    \centering
    \includegraphics[width=1\linewidth]{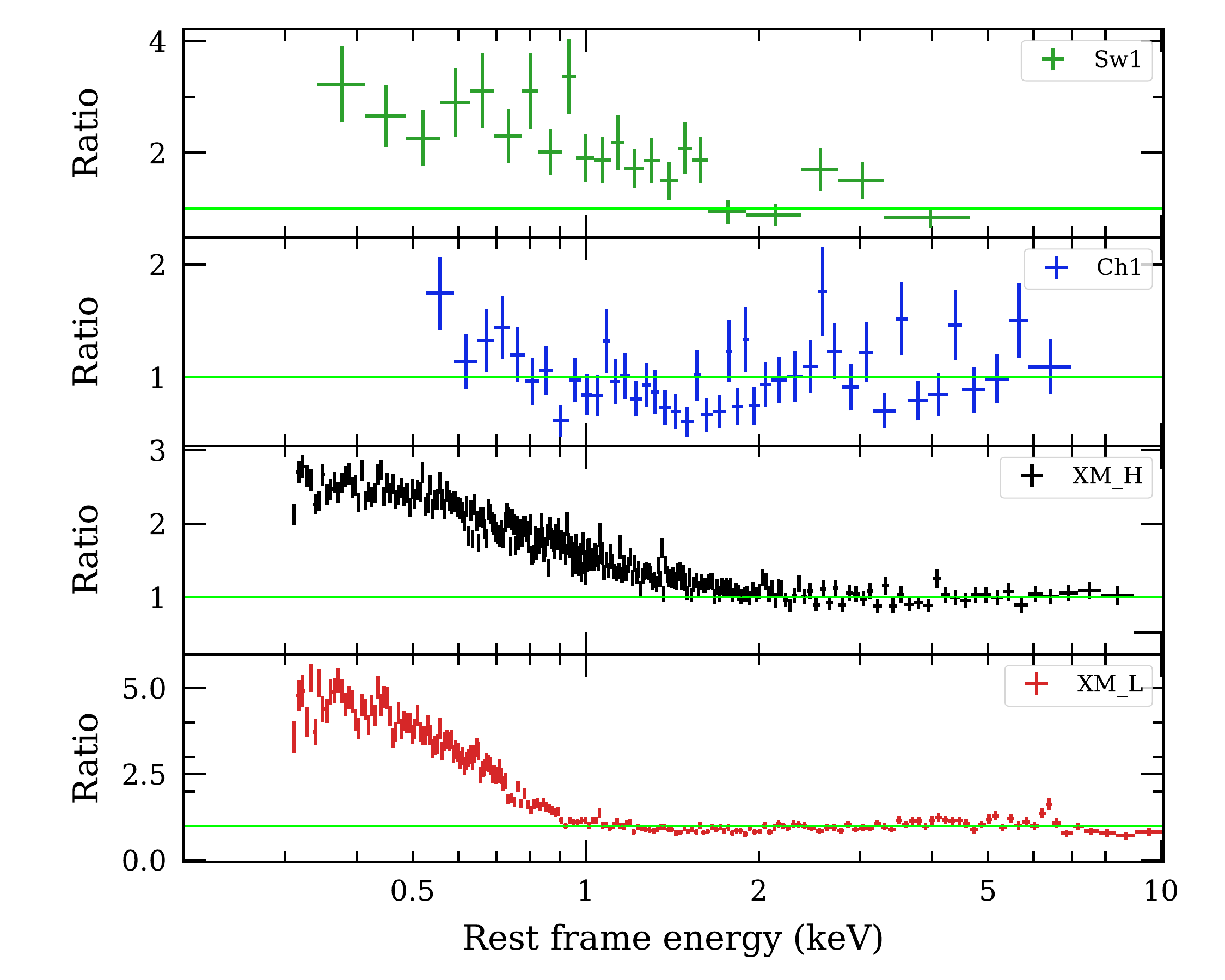}
    \caption{Ratio of the Sw1, Ch1, XM\_H, and XM\_L spectra of Mrk~382
relative to a Galactic-absorbed power-law model fitted in the hard
X-ray band (2--10 keV). A prominent soft X-ray excess below
$\sim2$ keV is clearly visible in all
spectra. For the multiple \textit{Swift} observations, only the Sw1
spectrum is shown as a representative example.}\label{fig:po}
\end{figure}

\section{Results} \label{sec:X-RAY AND THE MUTILWAVELENGTH PROPERTIE}
\subsection{X-ray Spectral Analysis}
\label{subsec:X-Ray spectra Analysis}
We performed spectral fitting using \texttt{XSPEC} (version 12.13; \citealt{Arnaud1996}). The spectra
were fitted over energy ranges chosen based on the signal-to-noise ratio
of each observation. The $\chi^2$ statistic was used for the {\xmm},
\textit{Swift}, \textit{Chandra}, and eROSITA spectra, while the Cash
statistic \citep{cash1979} was adopted for the low-count
\textit{Swift} observations (Sw3, Sw5, Sw7, Sw9, and Sw10). All quoted
uncertainties correspond to the 90\% confidence level for one interesting
parameter ($\Delta\chi^2 = 2.71$). Galactic absorption was modeled using
\texttt{tbabs}, with the neutral hydrogen column density fixed at
$N_{\rm H}=4.69\times10^{20}\ {\rm cm^{-2}}$, as derived from the
HEASARC online $N_{\rm H}$ tool \citep{2016HI4PI}.

\begin{figure}
    \includegraphics[width=1\linewidth]{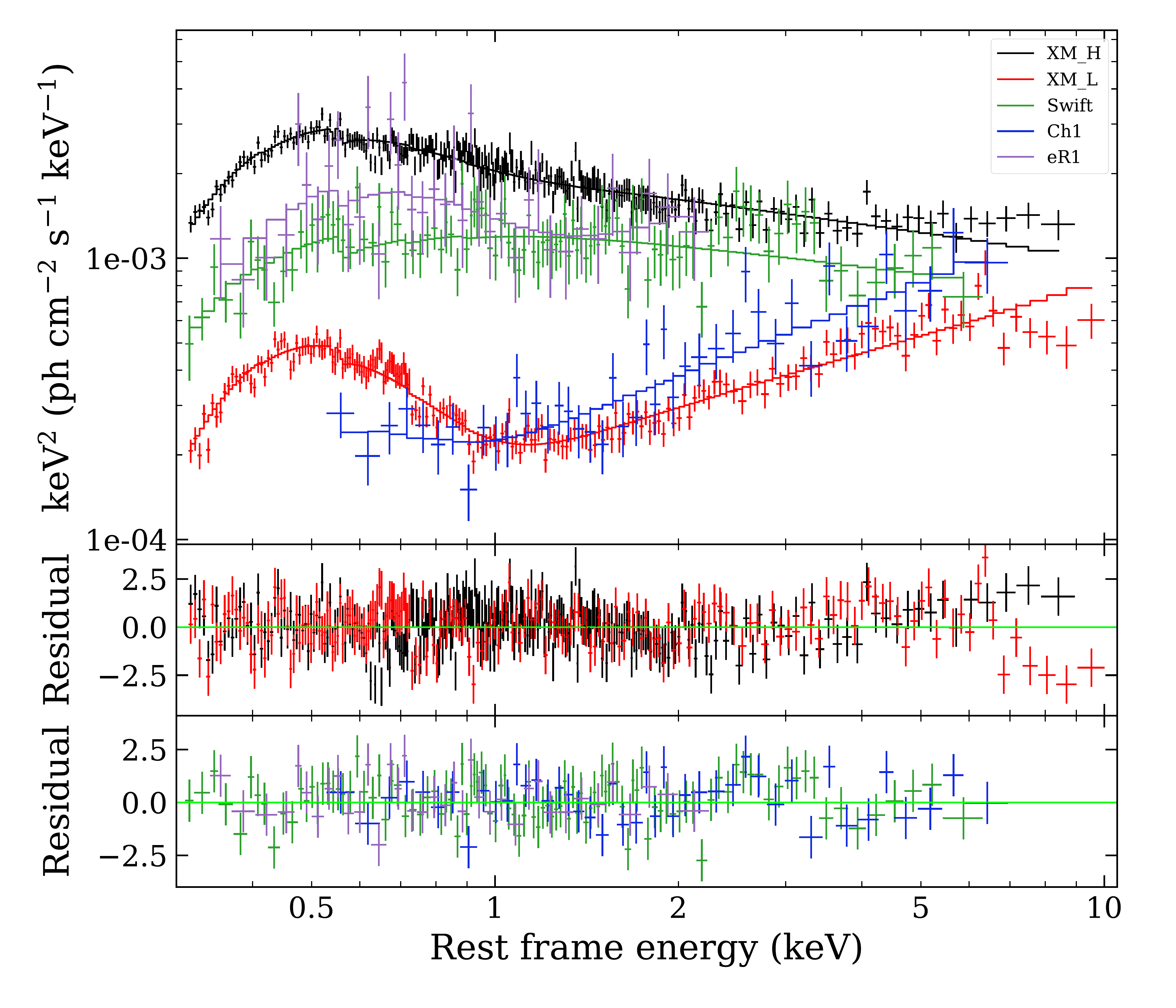}
    \caption{Unfolded X-ray spectra of Mrk~382 from the XM\_H (black),
    XM\_L (red), combined \textit{Swift} (green), Ch1 (blue), and eR1
    (purple) observations, fitted with a power-law plus blackbody model
    modified by Galactic absorption. For clarity, only the stacked
    \textit{Swift} spectrum is shown. The lower panel shows the
    normalized residuals, defined as
    $(\mathrm{data}-\mathrm{model})/\mathrm{error}$.}
    \label{fig:po_zbb}
\end{figure}

We first fitted the hard X-ray band (2--10 keV) to characterize the
underlying continuum. Extrapolating the model to lower energies reveals
clear residuals below $\sim2$ keV, indicating the presence of a soft
X-ray excess (Figure~\ref{fig:po}). To account for this component, we
added a blackbody, adopting a \texttt{tbabs*(zpowerlw+zbbody)} model.
This addition significantly improves the fits for most observations
(Figure~\ref{fig:po_zbb}), and the resulting parameters are listed in
Table~\ref{tab:po_zbb}.

\begin{deluxetable*}{cccccccccc}
\tabletypesize{\scriptsize}
\tablecaption{Spectral fitting results for the X-ray observations of Mrk~382 using a Galactic-absorbed power-law plus blackbody model.}
\tablenum{2}
\label{tab:po_zbb}
\tablewidth{0pt}
\tablehead{
\colhead{Date} &
\colhead{Abbr.} &
\colhead{$\Gamma$} &
\colhead{$kT$} &
\colhead{$N_{\rm bb}$} &
\colhead{$\chi^2$/dof} &
\colhead{C-stat/dof} &
\colhead{$F_{\rm 0.3-2}$} &
\colhead{$F_{\rm 2-10}$} &
\colhead{Net}  \\
\colhead{} &
\colhead{} &
\colhead{} &
\colhead{(keV)} &
\colhead{($10^{-6}$)} &
\colhead{} &
\colhead{} &
\colhead{($10^{-12}$)} &
\colhead{($10^{-12}$)} &
\colhead{counts}
}
\startdata
2009-02-26 & Sw1 &
$1.88^{+0.39}_{-0.56}$ &
$0.18^{+0.06}_{-0.04}$ &
$29^{+20}_{-24}$ &
21/18 &
$\cdots$ &
$5.7^{+0.6}_{-0.5}$ &
$3.1^{+0.7}_{-0.6}$ &
472 \\
2009-08-31 & Sw2 &
$2.24^{+0.14}_{-0.13}$ &
$0.10^{f}$ &
$<27$ &
20/19 &
$\cdots$ &
$7.0\pm0.7$ &
$3.9^{+0.8}_{-0.7}$ &
466 \\
2010-12-06 & Ch1 &
$1.19^{+0.14}_{-0.15}$ &
$0.10^{+0.04}_{-0.04}$ &
$7^{+21}_{-4}$ &
45/40 &
$\cdots$ &
$0.94\pm0.10$ &
$2.1^{+0.3}_{-0.3}$ &
1219 \\
2011-08-31 & Sw3 &
$2.37^{+0.61}_{-0.58}$ &
$0.10^{f}$ &
$90^{+60}_{-60}$ &
$\cdots$ &
12/7 &
$5.4^{+2.8}_{-2.0}$ &
$2.4^{+3.2}_{-1.5}$ &
21 \\
2011-09-01 & Sw4 &
$2.12\pm0.10$ &
$0.10^{f}$ &
$<16$ &
22/29 &
$\cdots$ &
$4.7\pm0.4$ &
$3.0^{+0.5}_{-0.5}$ &
691 \\
2011-11-02 & XM\_H &
$2.30\pm0.04$ &
$0.114\pm0.004$ &
$45\pm4$ &
290/256 &
$\cdots$ &
$9.1\pm0.1$ &
$3.30^{+0.08}_{-0.08}$ &
46007 \\
2019-10-29 & Sw5 &
$2.21\pm0.31$ &
$0.10^{f}$ &
$18\pm10$ &
$\cdots$ &
11/6 &
$1.4^{+0.4}_{-0.3}$ &
$0.8^{+0.5}_{-0.3}$ &
79 \\
2019-10-30 & XM\_L &
$1.38\pm0.03$ &
$0.100\pm0.002$ &
$18.0\pm0.1$ & 243/125 &
$\cdots$ &
$1.40^{+0.04}_{-0.03}$ &
$1.33\pm0.04$ &
25775 \\
2020-04-18 & eR1 &
$2.35^{+0.22}_{-0.86}$ &
$0.10^{f}$ &
$<46$ &
40/36 &
$\cdots$ &
$6.6^{+0.7}_{-0.6}$ &
$2.1^{+0.7}_{-0.6}$ &
172 \\
2021-01-01 & Sw6 &
$2.28^{+0.16}_{-0.15}$ &
$0.10^{f}$ &
$<17$ &
16/15 &
$\cdots$ &
$4.3\pm0.5$ &
$1.9^{+0.5}_{-0.4}$ &
368 \\
2021-01-03 & Sw7 &
$2.06\pm0.16$ &
$0.10^{f}$ &
$12^{+9}_{-11}$ &
70/45 &
$\cdots$ &
$2.5\pm0.3$ &
$2.2^{+0.6}_{-0.5}$ &
262 \\
2025-04-24 & Sw8 &
$2.37^{+0.23}_{-0.31}$ &
$0.10^{f}$ &
$11^{+30}_{-11}$ &
16/16 &
$\cdots$ &
$7.2^{+0.7}_{-0.8}$ &
$2.5^{+0.6}_{-0.5}$ &
399 \\
2025-05-22 & Sw9 &
$1.89^{+0.43}_{-0.42}$ &
$0.10^{f}$ &
$22\pm22$ &
$\cdots$ &
21/12 &
$4.7^{+0.8}_{-0.7}$ &
$2.7^{+1.0}_{-0.8}$ &
152 \\
2025-09-13 & Sw10 &
$2.15^{+0.29}_{-0.26}$ &
$0.10^{f}$ &
$14^{+10}_{-9}$ &
$\cdots$ &
18/12 &
$3.3^{+0.6}_{-0.4}$ &
$1.2^{+0.5}_{-0.4}$ &
148 \\
2025-09-29 & Sw11 &
$1.98\pm0.23$ &
$0.10^{f}$ &
$<17$ &
$\cdots$ &
$\cdots$ &
$5.0^{+0.8}_{-0.9}$ &
$4.4^{+1.7}_{-1.3}$ &
132 \\
2025-10-11 & Sw12 &
$1.66^{+0.53}_{-0.59}$ &
$0.10^{f}$ &
$13^{+22}_{-13}$ &
$\cdots$ &
3/5 &
$3.5^{+0.9}_{-0.7}$ &
$3.1^{+2.0}_{-1.3}$ &
77 \\
2025-12-11 & Sw13 &
$2.21^{+0.46}_{-0.65}$ &
$0.10^{f}$ &
$22^{+42}_{-22}$ &
$\cdots$ &
15/11 &
$6.3^{+1.2}_{-1.0}$ &
$2.4^{+1.1}_{-0.8}$ &
140 \\
\enddata
\tablecomments{
Columns (1)--(2): observation date and abbreviation.
Column (3): photon index.
Column (4): blackbody temperature.
Column (5): blackbody normalization.
Columns (6) and (7): fit statistic ($\chi^2$ or Cash statistic) and corresponding degrees of freedom.
Columns (8) and (9): observed 0.3--2.0 keV and 2--10 keV fluxes in units of $10^{-12}\ {\rm erg\ cm^{-2}\ s^{-1}}$.
Column (10): net source counts.
Parameters marked with the superscript $f$ were fixed during the fitting because they could not be well constrained.
}
\end{deluxetable*}

The photon indices derived from the \textit{Swift} observations span a
range of $\Gamma \sim 1.7$--2.4, although the relatively large
uncertainties prevent a clear identification of spectral evolution.
In contrast, the higher signal-to-noise spectra exhibit more pronounced
variations. The photon index changes from $\Gamma = 2.32\pm0.04$ in the
high-flux state (XM\_H) to $\Gamma = 1.39\pm0.06$ in the low-flux state
(XM\_L), and $\Gamma = 1.19^{+0.14}_{-0.15}$ in the \textit{Chandra}
observation. This trend indicates a clear spectral hardening as the
source flux decreases, consistent with a harder-when-fainter behavior.
The 0.3--2.0 keV flux exhibits substantial variability across epochs,
ranging from $\sim0.9\times10^{-12}$ to $\sim9\times10^{-12}\ \rm erg\
cm^{-2}\ s^{-1}$ (Table~\ref{tab:po_zbb}), corresponding to a variability
amplitude of nearly an order of magnitude. Notably, the XM\_L and Ch1
observations correspond to relatively low-flux states, while XM\_H
represents a high-flux state, consistent with the spectral changes
described above.

In addition to the continuum components, the residual spectra reveal
complex spectral features across the X-ray band (see
Figure~\ref{fig:po_zbb}), indicating the presence of additional
spectral components beyond the adopted phenomenological model. In
particular, low-energy residuals below $\sim$1 keV are present in both
the XM\_H and XM\_L observations, which may be associated with
additional photoionized absorption or emission components, as commonly
reported in Seyfert galaxies (e.g., \citealt{komossa2000}).

Furthermore, the XM\_L spectrum exhibits notable residuals in the
Fe-K band, including an emission feature around $\sim$6.3--6.4 keV and
absorption-like deviations at higher energies ($\sim$7--9 keV). Such
complex Fe-K spectral features have been interpreted as arising from
either additional ionized absorption components or relativistic
reflection features (e.g., \citealt{middei2023}).

Recently, \citet{xu2025} demonstrated that the broadband
X-ray spectrum of Mrk~382, combining \xmm\ and \textit{NuSTAR}
observations, can be well described by a relativistically blurred
reflection model, which naturally accounts for both the soft X-ray excess
and the Compton hump. Motivated by this result, we first applied the
\texttt{relxill} model to the XM\_H and XM\_L spectra to investigate
whether ionized reflection alone can reproduce the observed spectral
features.

Following the setup adopted by \citet{xu2025}, we employed a broken
power-law emissivity profile, with the outer emissivity index fixed at 3
and the break radius set to $6\,R_{\rm g}$. The inner disk radius was
fixed at the innermost stable circular orbit (ISCO), while the outer
radius was fixed at $500\,R_{\rm g}$. The high-energy cutoff was fixed
at 300 keV. The black hole spin, disk inclination, and iron abundance
were tied between the two observations, whereas the photon index,
ionization parameter, and normalization were allowed to vary.

Although the reflection model does not provide a statistically significant
improvement over the phenomenological model, it offers a physically
motivated description of the broad-band spectral shape. In particular,
the relativistic reflection component can naturally account for the soft
X-ray excess and the spectral curvature above $\sim3$ keV. The reflection fraction shows a strong dependence on the
flux state, increasing from $R_{\rm refl}\sim4$ in the high-flux state
(XM\_H) to $R_{\rm refl}\sim34$ in the low-flux state (XM\_L). Such a large reflection fraction indicates a
reflection-dominated spectrum during the low state. However, systematic
deviations below $\sim1$ keV remain, and in addition, the emission feature near the neutral Fe K$\alpha$
energy ($\sim6.4$ keV) and absorption-like deviations at
$\sim7$--$9$ keV in the low-flux state can not be well characterized by the reflecion model.

We then introduce an additional redshifted Gaussian component 
(\texttt{zgauss}) to the XM\_L spectrum  to model the emission feature around $\sim$6.3 keV,
which is not fully accounted for by the relativistic reflection model.
The redshift of the Gaussian component is fixed to the systemic
redshift of Mrk~382. The line centroid energy is measured to be
$E = 6.35^{+0.06}_{-0.04}$ keV in the rest frame, with an equivalent
width of $\mathrm{EW}=330$ eV. The inclusion of this component
significantly improves the fit ($\Delta\chi^2=18$ for $\Delta\nu=2$),
indicating a statistically significant emission feature. The line width
is not well constrained and is consistent with being unresolved;
therefore, it is fixed to the instrumental resolution of 150 eV. The
centroid energy is consistent with Fe K$\alpha$ emission from
low-ionization material, suggesting an origin in distant, cold
reflecting gas.

To investigate whether the narrow Fe K$\alpha$ component detected in the
XM\_L spectrum is also present in the XM\_H state, we added a narrow
Gaussian component with the centroid energy fixed at $E=6.35$ keV and
the line width fixed at 150 eV. The inclusion of this component does not improve the fit, indicating that
the narrow Fe K$\alpha$ emission is not required by the data in the
high-flux state. This may be due to dilution of the line equivalent
width by the stronger primary continuum during the high-flux state.

Considering these spectral features, we employed a combination of physically motivated models to describe the different processes contributing to the observed X-ray spectra. The primary X-ray continuum and relativistic reflection from the inner accretion disk were modeled with \texttt{relxill}. We additionally included an \texttt{xillver} component to account for distant reflection and the associated narrow Fe K$\alpha$ emission. The photon index, iron abundance, high-energy cutoff, and inclination of \texttt{xillver} were tied to the corresponding parameters of \texttt{relxill}, while its reflection fraction was fixed at $-1$ to retain only the reflected emission. The ionization parameter and normalization of the \texttt{xillver} component were allowed to vary independently.

We further included an ionized absorber, implemented as a precomputed
XSTAR multiplicative table model \citep{Kallman2001}, to characterize the
remaining soft X-ray absorption features in both the XM\_H and XM\_L
spectra. The photon index of the ionizing continuum in the XSTAR model was
tied to that of the corresponding \texttt{relxill} continuum for each flux
state, while the turbulent velocity was fixed at
$200~{\rm km~s^{-1}}$. Solar abundances were adopted for all elements. The column
density ($N_{\rm H}$) and ionization parameter ($\log\xi$) of the
photoionized absorber were allowed to vary during the spectral fitting. We initially allowed the absorber redshift to vary, but found that
it could not be robustly constrained by the spectra. We therefore fixed
the absorber redshift at the systemic redshift of Mrk~382
($z=0.0332$) in the final fits.

\begin{figure}
      \includegraphics[width=1\linewidth]{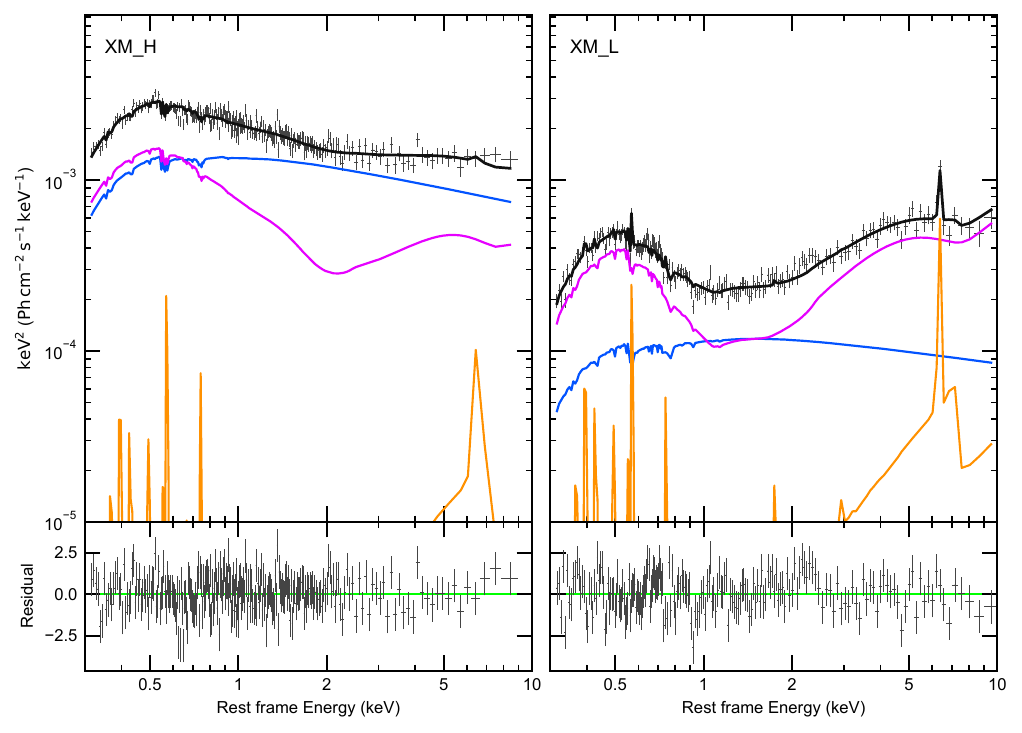}
      \caption{Unfolded spectra of the XM\_H (left) and XM\_L (right)
observations of Mrk~382 fitted with a model consisting of Galactic
absorption, a photoionized warm absorber, relativistic disk reflection,
and distant reflection. The individual model components and the total
best-fitting model are shown separately. The lower panels show the
normalized residuals, defined as (data$-$model)/error.}\label{fig:bestfit}
\end{figure}

The resulting models provide an adequate description of both
XMM-Newton spectra (Figure~\ref{fig:bestfit}), where the unfolded
spectra, individual model components, total model, and fitting
residuals are shown. The photoionized absorber exhibits different
ionization states between the two flux states. In XM\_H, the absorber
favors a relatively high ionization state, with
$\log\xi_{\rm warm}=3.4^{+0.4}_{-1.8}$, although the solution is poorly
constrained and the column density reaches the upper boundary of the
XSTAR grid ($N_{\rm H,warm}=10^{23}\ {\rm cm^{-2}}$). In contrast, the
absorber in XM\_L is better constrained, with
$\log\xi_{\rm warm}=1.60^{+0.02}_{-0.06}$ and
$N_{\rm H,warm}\simeq3\times10^{21}\ {\rm cm^{-2}}$.
Thus, although the absorber parameters in XM\_H should not be
interpreted quantitatively, the fits are consistent with a more highly
ionized absorber in the high-flux state. Such a change would
qualitatively make the absorber more transparent in XM\_H, whereas the
lower ionization state in XM\_L produces more prominent absorption
features in the soft X-ray spectrum.

The spectral decomposition in Figure~\ref{fig:bestfit} further
illustrates a pronounced change in the relative contributions of the
direct and reflected emission between the two X-ray states. In XM\_H,
both the directly observed coronal continuum and relativistic disk
reflection contribute substantially to the spectrum. In XM\_L,
however, the direct continuum is strongly suppressed relative to the
relativistic reflection component, which dominates much of the
observed X-ray emission and contributes strongly to both the soft
excess and the broad Fe-K spectral curvature. Correspondingly, the
disk ionization parameter decreases from
$\log\xi_{\rm disk}=2.6^{+0.1}_{-0.3}$ in XM\_H to
$\log\xi_{\rm disk}=1.6\pm0.1$ in XM\_L, while the reflection fraction
increases from $R_{\rm refl}=4.4^{+1.8}_{-1.1}$ to
$R_{\rm refl}=30.2^{+6.7}_{-6.4}$. Within the adopted model, this
spectral decomposition indicates that the XM\_L spectrum is strongly
reflection dominated, broadly consistent with the relativistic
reflection interpretation proposed by \citet{xu2025}. In a
light-bending scenario, such behavior may be associated with a more
compact corona and/or a decrease in its characteristic height above
the disk, such that a larger fraction of the coronal emission
illuminates the inner disk rather than reaching the observer directly.

The more complete spectral model adopted here additionally includes
distant reflection to self-consistently account for the narrow
Fe K$\alpha$ emission and its associated reflection continuum. The
normalization of this component is well constrained in XM\_L, whereas
it is consistent with zero within the uncertainties in XM\_H. The
ionization parameter of the distant \texttt{xillver} reflector was
initially allowed to vary but consistently reached the lower boundary
of the model; we therefore fixed it at $\log\xi=0$, corresponding to a
nearly neutral reflector. Thus, although photoionized absorption and
distant reflection contribute to the detailed spectral shape, the
strong increase in the relative contribution of relativistic
reflection toward the low-flux state remains a prominent feature of
the spectral decomposition. The residuals shown in
Figure~\ref{fig:bestfit} demonstrate that the combined model provides
an adequate description of both spectra, with
$\chi^2/\nu=254/252$ and $186/189$ for XM\_H and XM\_L, respectively,
while accounting for the principal spectral structures in the soft
X-ray and Fe-K bands.

A weak absorption-like residual remains near
$6.5$ keV in the observed frame ($\sim6.7$ keV in the source rest
frame) in the XM\_L spectrum. We modeled this feature with a narrow
\texttt{gabs} component, obtaining a line centroid energy of
$E_{\rm abs}=6.5^{+0.3}_{-0.1}$ keV, with the line width fixed at
$\sigma=0.064$ keV. The inclusion of \texttt{gabs} improves the fit
by $\Delta\chi^2=4$ for two additional free parameters. Given the
modest improvement in the fit, we regard this absorption feature as
tentative and do not claim a statistically significant line
detection.

\begin{deluxetable}{lccc}
\tabletypesize{\scriptsize}
\tablecaption{Best-fit model parameters for the XMM-Newton/EPIC-pn
XM\_H and XM\_L spectra. Uncertainties correspond to 90\% confidence
intervals for one parameter of interest ($\Delta\chi^2=2.706$).
Superscripts $f$, $*$, and $p$ denote fixed, tied, and
hard-limit-pegged parameters, respectively, while $\dagger$ indicates
a non-monotonic XSPEC error profile.}
\tablenum{3}
\label{tab:xspec_90pct}
\tablewidth{0pt}
\tablehead{
Parameter &
Unit &
\colhead{XM\_H} &
\colhead{XM\_L}
}
\startdata
\hline
\multicolumn{4}{c}{\texttt{warm absorber}} \\
\hline
$N_{\rm H,warm}$ & $10^{21}\,\mathrm{cm^{-2}}$ &
$100^{p}$ &
$3.0^{+0.4}_{-0.4}$ \\
$\log\xi_{\rm warm}$ & $\mathrm{erg\,cm\,s^{-1}}$ &
$3.4_{-1.8}^{+0.4,p}$ &
$1.60_{-0.05}^{+0.02}$ \\
$z_{\rm warm}$ & -- &
$0.0332^{f}$ &
$0.0332^{f}$ \\
\hline
\multicolumn{4}{c}{\texttt{relxill}} \\
\hline
$q_{\rm in}$ & -- &
$10.0_{-0.4}^{\,p}$ &
$9.5^\dagger$ \\
$a_{*}$ & -- &
$0.998^{f}$ &
$0.998^{f}$ \\
Inclination & -- &
$62^{f}$ &
$62^{f}$ \\
$\Gamma$ & -- &
$2.33_{-0.04}^{+0.04}$ &
$2.21_{-0.02}^{+0.01}$ \\
$\log\xi$ & $\mathrm{erg\,cm\,s^{-1}}$ &
$2.6_{-0.3}^{+0.1}$ &
$1.6\pm0.1$ \\
$R_{\rm refl}$ & -- &
$4.4_{-1.1}^{+1.8}$ &
$30.2_{-6.4}^{+6.7}$ \\
Norm & $10^{-6}$ &
$19.8_{-1.2}^{+1.6}$ &
$2.8_{-0.1}^{+0.3}$ \\
\hline
\multicolumn{4}{c}{\texttt{xillver}} \\
\hline
Norm & $10^{-6}$ &
$8.2_{-8.2}^{+11}$ &
$8.2_{-0.3}^{+0.3}$ \\
\hline
\multicolumn{4}{c}{Fit statistics} \\
\hline
$\chi^2/\nu$ & -- &
$254/252$ &
$187/189$ \\
\hline
\enddata
\end{deluxetable}

\subsection{X-ray and multi-wavelength Variability}
\label{sec:XRV}

\begin{figure}
      \flushleft
      \includegraphics[width=1\linewidth]{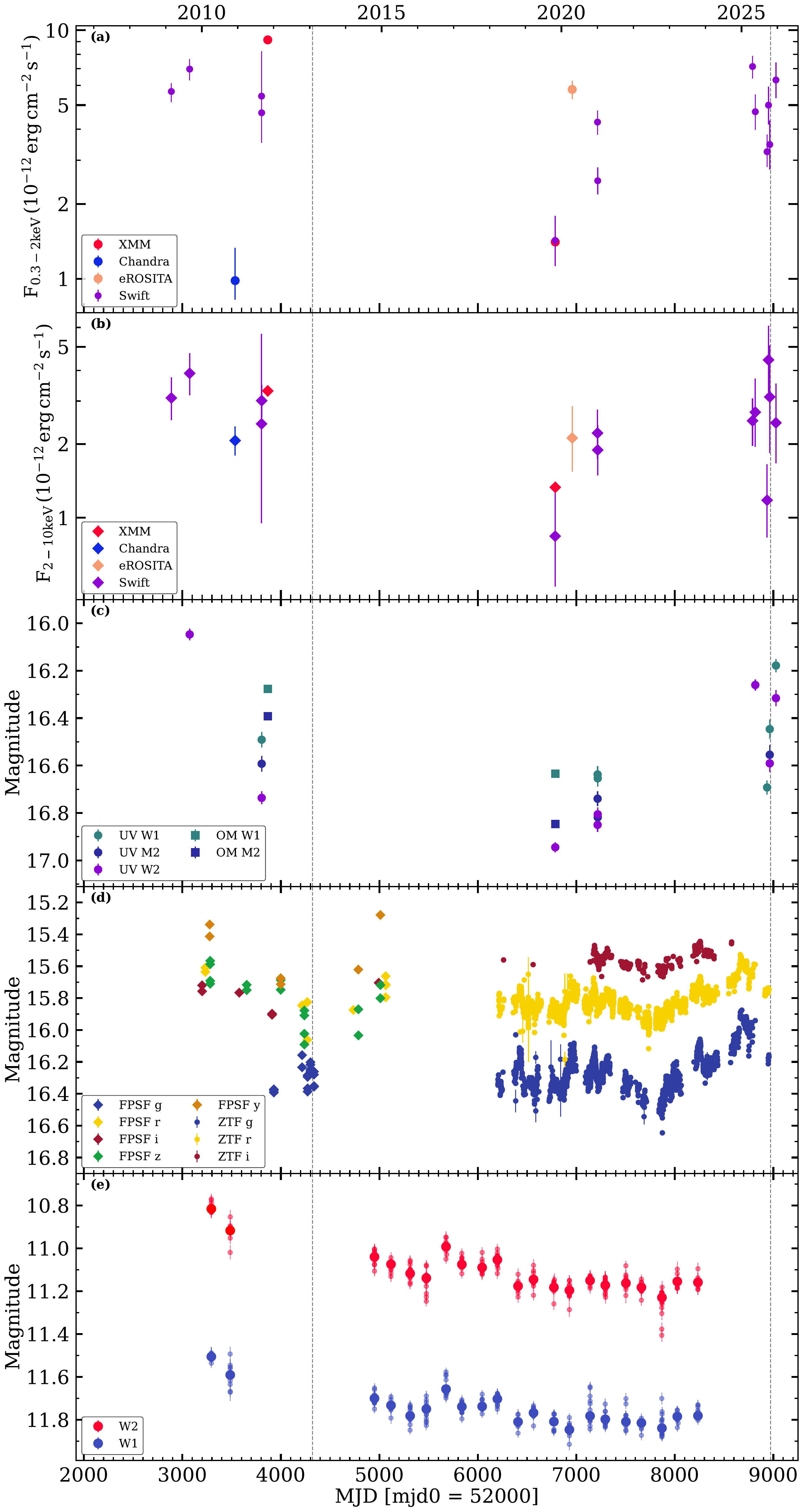}
      \caption{Light curves of the source in multiple bands: (a) the 0.3–2 keV soft X-ray flux; (b) the 2-10 keV hard X-ray flux;(c) the {\xmm}-OM and \textit{Swift}-UVOT W1, W2, and M2 bands; (d) the Pan-STARRS $g$, $r$, $i$, $z$, and $y$ bands together with the ZTF $g$, $r$, and $i$ bands; and (e) the NEOWISE W1 (3.4 $\mu$m) and W2 bands. The gray dotted vertical lines indicate the epochs of the optical spectroscopic observations. In panel (e), we also show the median magnitude for each observing season.}\label{fig:multilight curve}
\end{figure}

The X-ray emission of Mrk~382 shows pronounced variability on
timescales of years, as shown in Figure~\ref{fig:multilight curve}.
The source undergoes multiple transitions between bright and faint
states. The earliest \textit{Swift}/XRT observations in 2009 indicate
that Mrk~382 was already in a relatively bright state. By the 2010
\textit{Chandra} observation (Ch1), the soft X-ray flux had declined
by a factor of several, placing the source in a comparatively faint
state. This was followed by a dramatic brightening to the
\textit{XMM-Newton} high-flux state (XM\_H) in 2011, during which the
0.3--2 keV flux increased by a factor of $\sim10$ relative to Ch1.
This brightening trend was also supported by the nearly contemporaneous
\textit{Swift}/XRT monitoring. The source later underwent a major
decline, decreasing by a factor of $\sim6$--7 to the low-flux state
(XM\_L) observed in 2019. More recent \textit{Swift}/XRT observations
reveal a renewed brightening phase, with flux levels recovering to
values comparable to, or intermediate between, the historical high-
and low-flux states. Given the cadence of the available monitoring,
some of these state transitions likely occurred on timescales shorter
than one year. In comparison with the soft X-ray band, the 2–10 keV flux of
Mrk~382 does not show large-amplitude variability.

The ultraviolet variability was investigated using data from the
UVOT instrument onboard \textit{Swift} and the OM instrument onboard
\textit{XMM-Newton}, as shown in Figure~\ref{fig:multilight curve}.
The UVOT/OM measurements were corrected for Galactic extinction using
$E(B-V)=0.13$ \citep{s_f2011} and a Milky Way extinction
law with $R_V=3.1$ \citep{fitzpatrick2019}. The ultraviolet
variability shows a clear connection with the long-term evolution of
the soft X-ray emission, with fading and re-brightening transitions
also present in the UV bands. In particular, the UVOT observations
indicate that the source was relatively faint during the X-ray weaker
states around 2019--2021 and became brighter again during the recent
X-ray re-brightening phase in 2025. A similar trend is seen in the OM
measurements, where the source was brighter during the
\textit{XMM-Newton} high-flux state (XM\_H) and fainter during the
low-flux state (XM\_L). The amplitude of the UV variability is nevertheless much smaller than
that observed in X-rays. The largest change is seen in the UVW2 band,
with a maximum variability amplitude of $\sim0.9$ mag, corresponding
to a flux variation by a factor of $\sim2.3$, substantially smaller
than the contemporaneous X-ray variability.

To extend the analysis to longer wavelengths, we constructed
multiwavelength light curves using archival optical and infrared data,
including Pan-STARRS, ZTF ($g$, $r$, and $i$), and
\textit{WISE}/NEOWISE (W1 and W2), as shown in
Figure~\ref{fig:multilight curve}. 
The ZTF light curves were obtained using forced photometry and
were further binned into 3-day intervals to improve the
signal-to-noise ratio. 
For Pan-STARRS, we retrieved single-epoch forced PSF photometry
from the Pan-STARRS Data Release 2 (DR2) catalog
\citep{Flewelling2020}, hosted by the Mikulski Archive for Space
Telescopes (MAST). The dataset used in this work is available
through
\dataset[DOI: 10.17909/s0zg-jx37]{https://doi.org/10.17909/s0zg-jx37}. The mid-infrared light curves from the
\textit{WISE}/NEOWISE W1 and W2 bands were binned into 7-day intervals,
and the mean magnitudes were adopted. In the optical band, the Pan-STARRS and ZTF measurements reveal clear
variability at the $\sim0.3$--0.5 mag level across the $g$, $r$, and
$i$ bands. In contrast,
the infrared \textit{WISE} bands exhibit only relatively small
variations over the same period and show a gradual long-term fading
trend, differing from the re-brightening behavior observed in the
X-ray, UV, and optical bands.

To quantify the variability across multi-wavelength bands, we calculate fractional variability amplitude, $\rm F_{var}$, which provides normalized estimate of variability relative to the mean flux \citep{Vaughan2003}, defined as: 
\begin{equation}
F_{\rm var} = \sqrt{\frac{\sigma^2_\text{XS}}{\bar{x}^2}}
\end{equation}
where  $\bar{x}$ is the mean flux and $\sigma^2_\text{XS}$ is the `excess variance' \citep{nandra1997,edelson2002}, which is defined as the observed variance after correcting for the contribution of measurement errors.  
\begin{equation}
\rm \sigma^2_\text{XS} = S^2-\overline{\sigma^2_{\rm err}}
\end{equation}
where $\rm S^2$ is the total variance of the light curve and the $\overline{\sigma^2_{\rm err}}$ is the mean square error.

\begin{figure}
    \centering
    \includegraphics[width=1\linewidth]{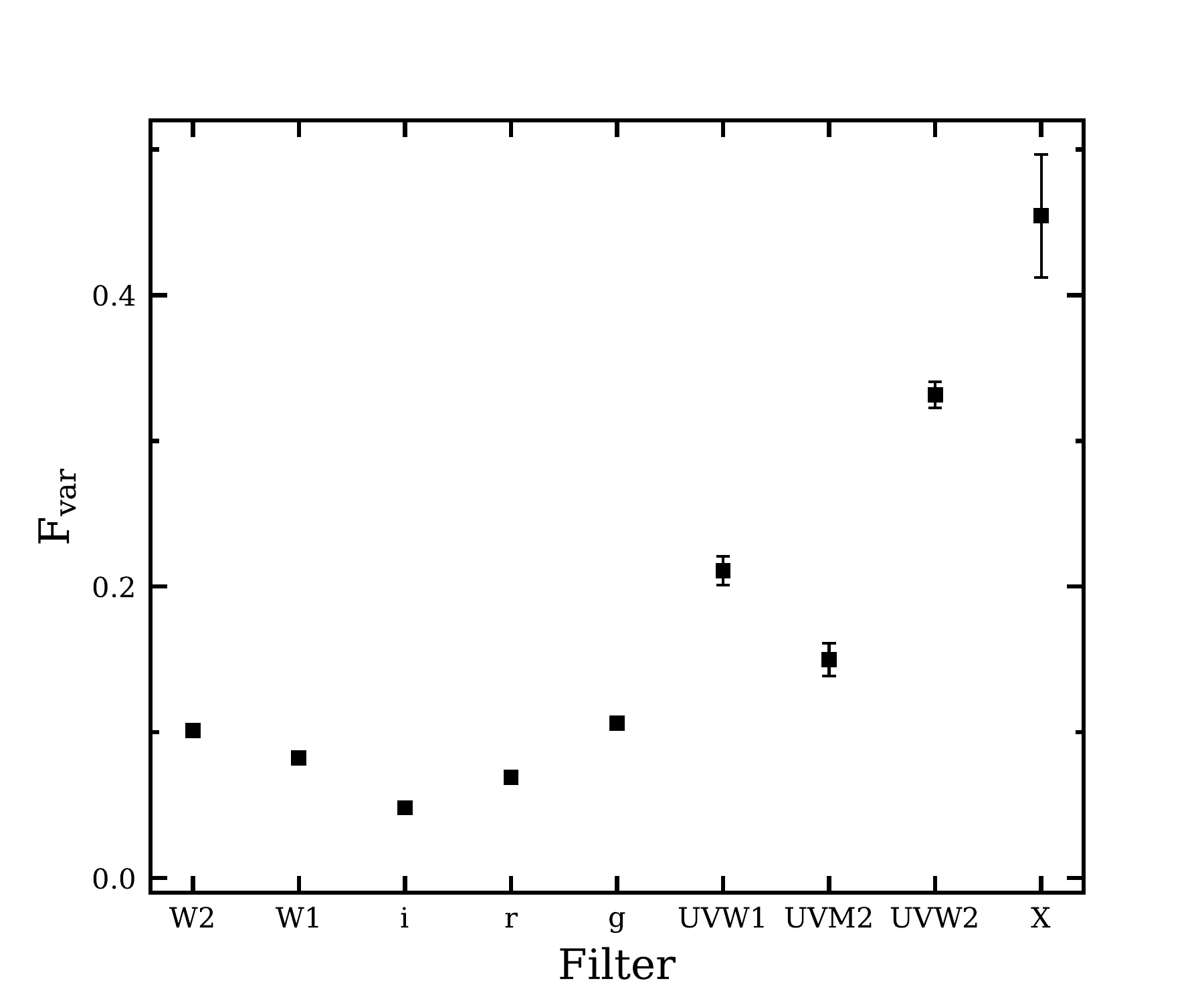}
    \caption{Fractional variability amplitude of Mrk 382 as a function of waveband,
from the infrared to the X-ray regime. The variability amplitudes are
derived from \textit{WISE} (W1 and W2), ZTF ($g$, $r$, and $i$),
\textit{XMM-Newton}/\textit{Swift} UV filters (UVW1, UVM2, and UVW2),
and X-ray observations from \textit{XMM-Newton}, \textit{Swift},
\textit{Chandra}, and \textit{eROSITA}. All error bars indicate $1\sigma$ uncertainties.}
    \label{fig:Fvar}
\end{figure}

The resulting fractional variability amplitudes in different bands  are shown in
Figure~\ref{fig:Fvar}. The uncertainty of $F_{\rm var}$ is estimated following \citet{Vaughan2003}, considering only the statistical uncertainties 
associated with the individual flux measurements.
This wavelength-dependent variability trend is further illustrated
by the fractional variability spectrum. The variability amplitude increases
systematically from the infrared and optical bands toward the
ultraviolet and X-ray regimes. The X-ray band exhibits the largest fractional variability,
with $F_{\rm var}\approx0.45$, while the UV bands show
intermediate variability amplitudes
($F_{\rm var}\approx0.15$--0.35). In contrast, the optical and infrared
bands display substantially smaller variability amplitudes. Such a
trend is naturally expected if the X-ray emission originates from a
compact and highly variable corona, whereas the longer-wavelength
emission is produced in progressively larger and more stable regions
of the accretion flow. The optical variability amplitudes in the $i$ and $r$ bands are
likely underestimated because Mrk~382 is a relatively nearby
Seyfert galaxy, and the observed optical emission may be
substantially diluted by host-galaxy starlight. The relatively lower variability amplitude measured in the UVM2 band
may also be partly affected by the sparse temporal sampling of the
available observations, which could lead to an underestimate of the
intrinsic variability amplitude in this band.

\subsection{Optical Spectroscopy}
We obtained a new optical spectrum of Mrk~382 on 2025 October 24 (JD~2460972)
using the 2.2\,m telescope at the Calar Alto Observatory
(Centro Astronómico Hispano en Andalucía; CAHA).
Two exposures of 1800 s each were taken with the Calar Alto Faint
Object Spectrograph (CAFOS), using the G-200 grism and a slit width of $3\farcs0$. The slit was oriented to include the same comparison star
used in the reverberation-mapping campaign of
\citet{hu2015}, enabling reliable relative flux
calibration. Data reduction and flux calibration were carried out
following the procedures described in
\citet{Hu2025}. The final spectrum covers an observed
wavelength range of $\sim4000$--8500~\AA\ with a dispersion of
4.47~\AA\ pixel$^{-1}$.

\begin{figure}
    \centering
    \includegraphics[angle=-90,width=\linewidth]{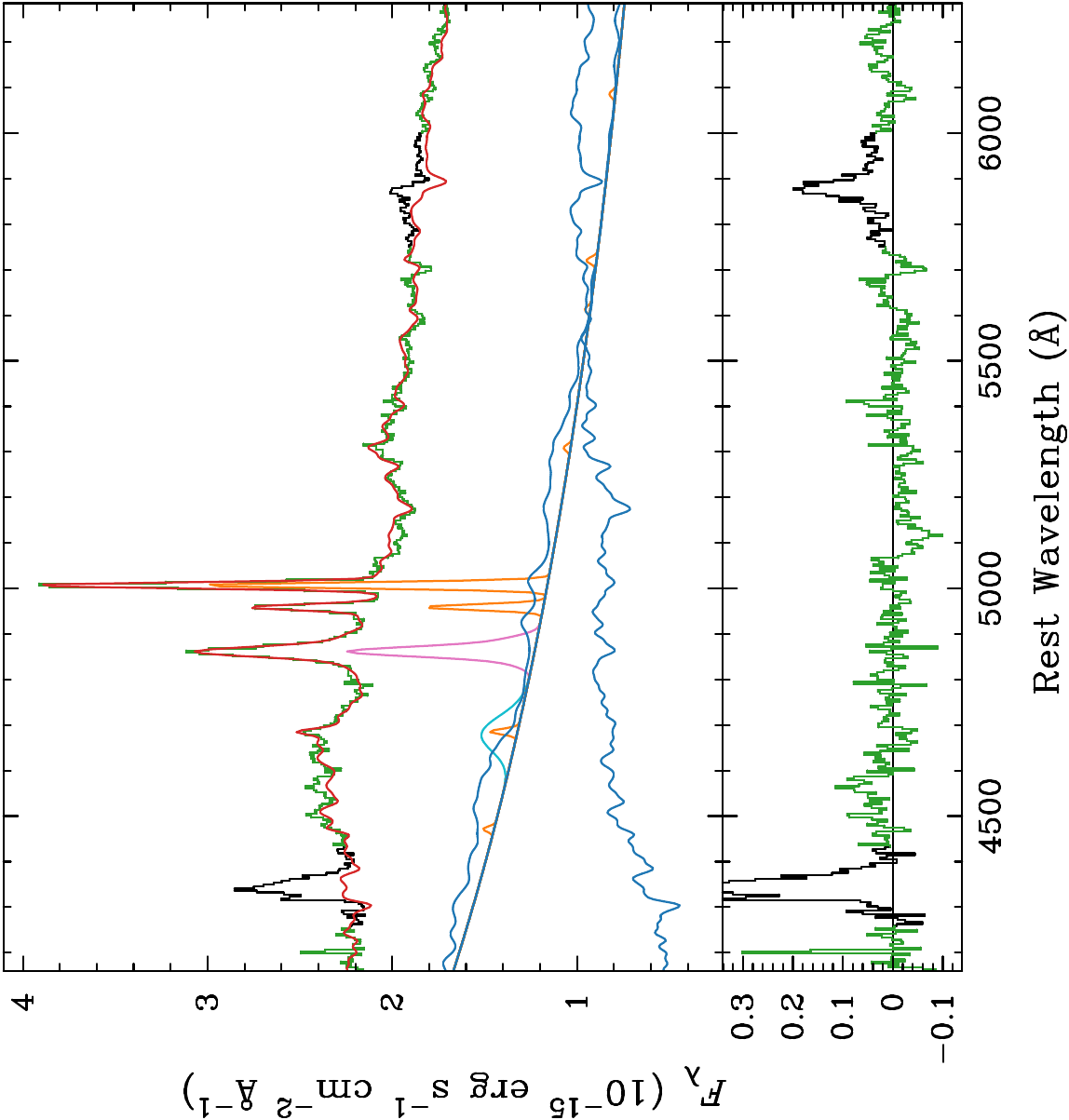}
    \caption{
    The new CAHA optical spectrum of Mrk~382 and its spectral decomposition.
Top: the observed spectrum within and outside the fitting windows
(shown in green and black, respectively). The best-fit model (red)
consists of the following components: the AGN power-law continuum,
Fe\,{\sc ii} pseudo-continuum, and host-galaxy starlight (all shown in
blue); the broad ${\rm H}\beta$ line (magenta); the broad He\,{\sc ii}
line (cyan); and the narrow emission-line components (orange).
Bottom: fit residuals.}
    \label{fig:caha}
\end{figure}

Figure~\ref{fig:caha} presents the Galactic-extinction-corrected,
rest-frame CAHA spectrum together with the spectral decomposition. We
modeled the spectrum following the method of
\citet{hu2015}, including an AGN power-law continuum, host
galaxy starlight, Fe\,II pseudo-continuum, and broad plus narrow
emission-line components. From the decomposition, we measure a host-subtracted AGN continuum flux density at 5100~\AA\ of
$F_{5100}=1.12\times10^{-15}\ {\rm erg\ s^{-1}\ cm^{-2}\ \AA^{-1}}$.
The broad H$\beta$ emission line has a flux of
$3.45\times10^{-14}\ {\rm erg\ s^{-1}\ cm^{-2}}$ and an intrinsic FWHM
of $1444\ {\rm km\ s^{-1}}$ after correcting for instrumental
broadening. Comparing to the
observations performed during years 2012--2013 (JD 2456224--2456421) in
\citet{hu2015} (see their Figures 1 and 4, and Table 2), both the
AGN continuum and H$\beta$ emission show no significant variabilities.

\begin{figure}
    \centering
    \includegraphics[width=\linewidth]{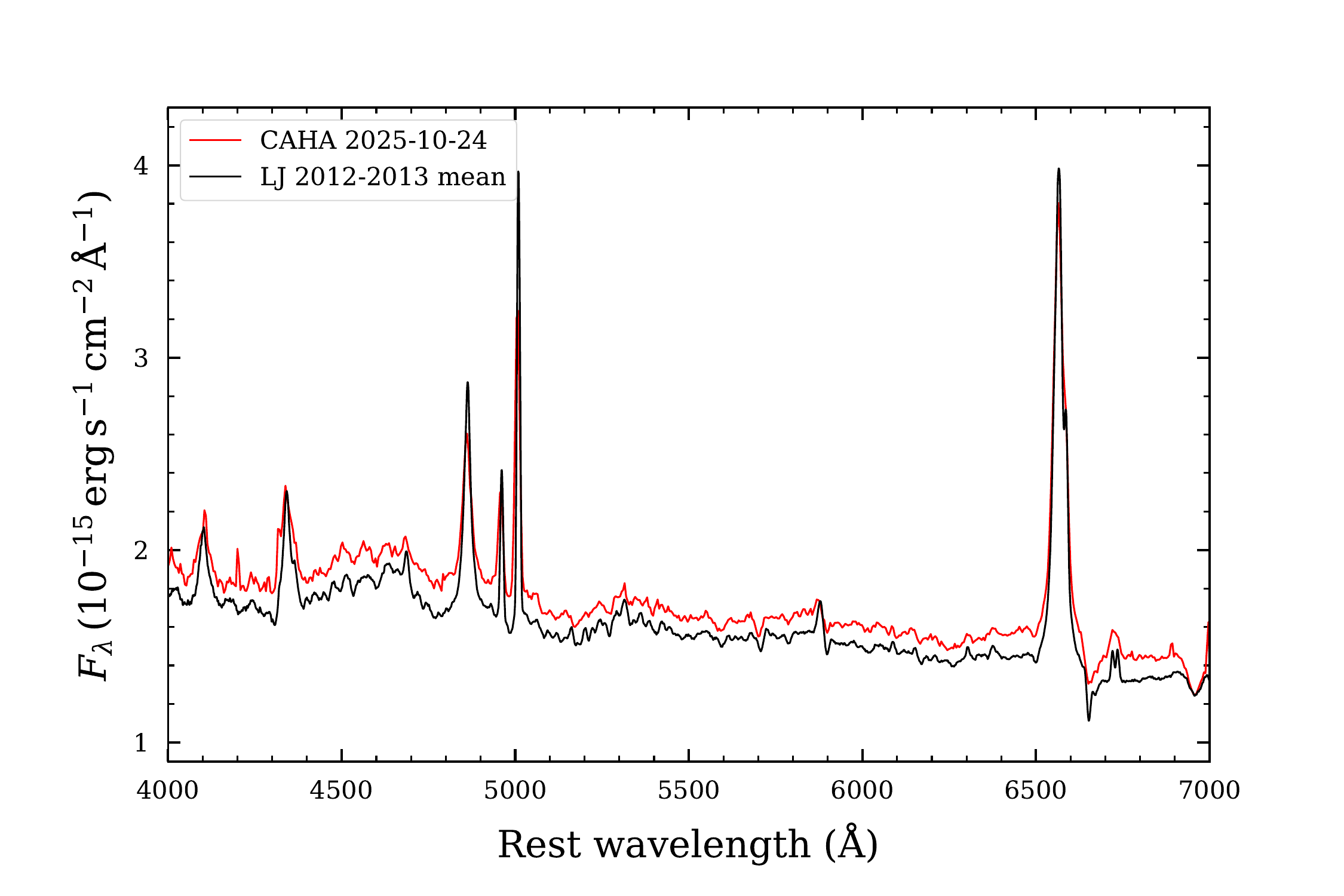}
    \caption{Optical spectra of Mrk~382 obtained in 2025 together with the mean spectrum from the 2012--2013 monitoring campaign by \citet{hu2015}.}
    \label{fig:multispec}
\end{figure}

For comparison, Figure~\ref{fig:multispec} shows the new CAHA spectrum
together with the Lijiang 2012--2013 mean spectrum
\citep{hu2015}. The overall spectral shape, continuum
level, and broad Balmer emission lines remain remarkably similar over
the $\sim12$ yr baseline. In particular, neither the AGN continuum near
5100~\AA\ nor the broad H$\beta$ line flux exhibits significant secular
changes within the measurement uncertainties. The broad-line profile
also appears stable between the two epochs. This long-term optical
stability contrasts with the pronounced X-ray variability observed over
the same general period. However, because no strictly simultaneous
X-ray observations are available for the 2013 Lijiang spectrum, we cannot
determine whether the two optical spectra correspond to identical or
different X-ray flux states. Therefore, the present comparison mainly
indicates that the optical broad-line region and continuum did not
undergo dramatic long-term changes, while any short-timescale response
to specific X-ray state transitions remains unconstrained.

\subsection{Broadband Spectral Energy Distribution}

\begin{figure}
    \centering
    \includegraphics[width=1\linewidth]{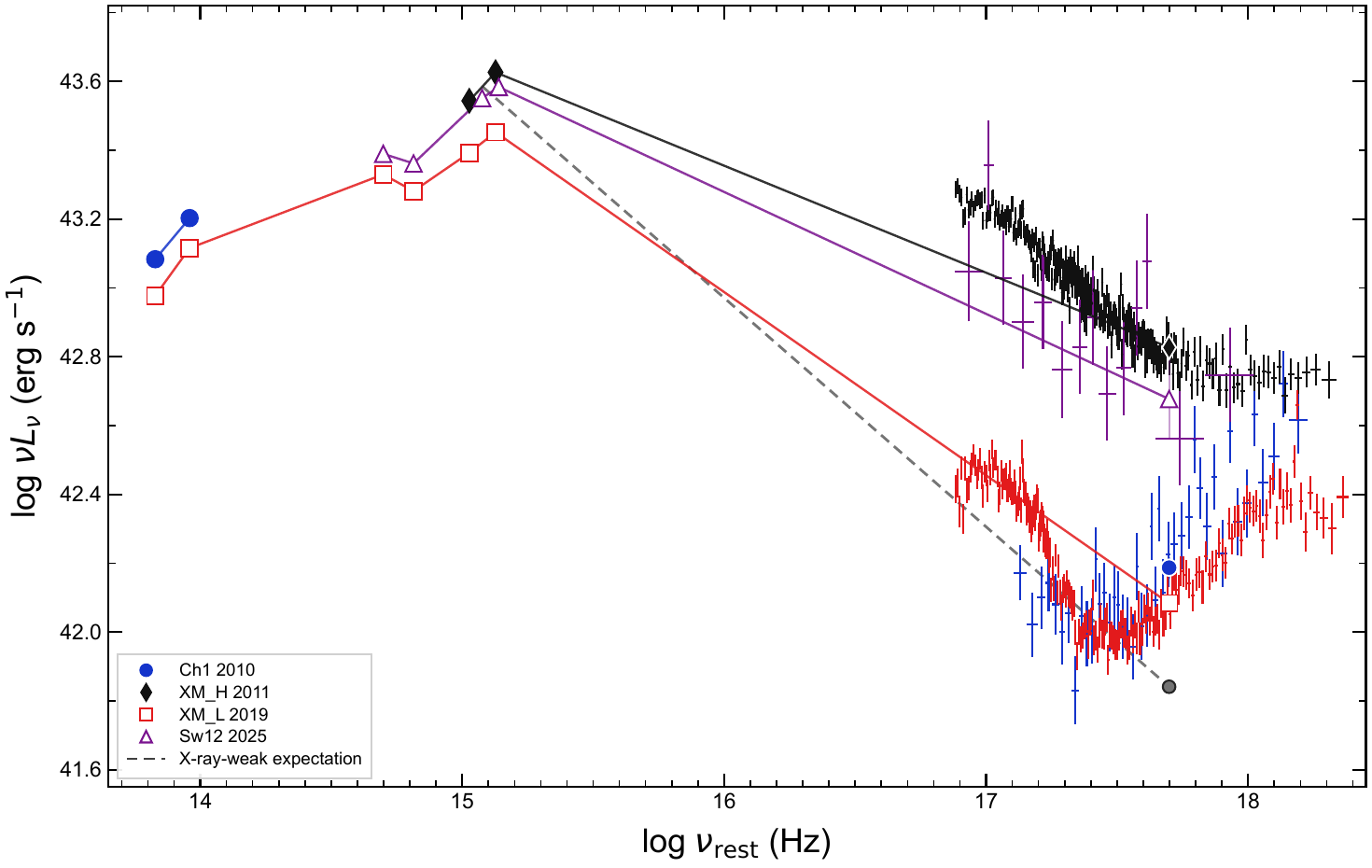}
    \caption{Rest-frame infrared-to-X-ray spectral energy distributions (SEDs) of Mrk~382 at four representative epochs. The infrared-to-UV data points are connected by solid lines to guide the eye, while the X-ray emission is shown as unfolded spectra derived from the blackbody plus power-law model. The gray point and dashed line
    indicate the expected 2\,keV luminosity of an
X-ray-weak AGN with the same UV luminosity as the XM\_H state,
assuming an X-ray weakness factor of $f_{\rm weak}\sim10$
relative to the standard
$\alpha_{\rm ox}$--$L_{2500\,\text{\AA}}$ relation of
\citet{steffen2006}.}
    \label{fig:sed}
\end{figure}

Figure~\ref{fig:sed} presents the rest-frame broadband spectral
energy distributions (SEDs) of Mrk~382, combining multiwavelength
data from the infrared to the X-ray bands for four representative
epochs: the 2010 \textit{Chandra} observation (Ch1), the 2011
\textit{XMM-Newton} high state (XM\_H), the 2019
\textit{XMM-Newton} low state (XM\_L), and the 2025
\textit{Swift} high state (Sw12).

The overall SED shape from the infrared through the optical/UV
bands remains broadly similar across different epochs, whereas the
largest variations occur in the X-ray regime. In particular, the
infrared ($\log \nu \lesssim 14.5$) and optical bands exhibit
only relatively modest variability, consistent with the long-term
light curves. The ultraviolet continuum associated with the big
blue bump shows intermediate variability amplitudes, while the
X-ray emission above $\log \nu \sim 17.5$ undergoes dramatic
changes. From the XM\_H state to the XM\_L state, the X-ray
luminosity decreases by nearly an order of magnitude, whereas the
optical/UV luminosity changes by only a factor of a few or less.
This behavior strongly suggests that the dominant driver of the
state transitions is associated primarily with the compact
X-ray-emitting corona rather than a global shutdown of the
accretion flow.

The persistence of a broadly similar optical/UV SED shape across
different epochs further indicates that the thermal emission from
the accretion disk remains largely intact even during the X-ray
faint state. Therefore, the low X-ray state is unlikely to be
caused by a dramatic decrease in the global mass accretion rate.
Instead, the observed SED evolution favors scenarios in which the
compact corona undergoes substantial changes in radiative
efficiency, geometry, or visibility. Possible explanations
include coronal collapse, vertical contraction of the corona with
enhanced gravitational light-bending effects, or partial
obscuration by inner-disk material.

A comparison with the expected 2\,keV luminosity inferred from
the UV luminosity of the XM\_H state further supports this
interpretation. As illustrated in Figure~\ref{fig:sed}, the gray
point marks the expected 2\,keV luminosity of an X-ray-weak AGN
with the same UV luminosity as the XM\_H state, assuming an
X-ray weakness factor of $f_{\rm weak}\sim10$ relative to the
standard $\alpha_{\rm ox}$--$L_{2500\,\text{\AA}}$ relation of
\citet{steffen2006}, corresponding to
$\Delta\alpha_{\rm ox}=-0.384$. Although the soft X-ray emission
of Mrk~382 is substantially suppressed during the low state, the
observed 2\,keV luminosity remains significantly higher than
that expected for a typical X-ray-weak AGN. This suggests that Mrk~382 does not enter a canonical
X-ray-weak state even during its faintest observed epoch,
despite the substantial suppression of the soft X-ray
emission.

\section{DISCUSSION} \label{sec:D}

\subsection{Origin of the Recurrent X-ray State Transitions}

Mrk~382 exhibits recurrent transitions between bright,
intermediate, and faint X-ray states on timescales of years,
with the soft X-ray flux varying by nearly an order of magnitude.
Such large-amplitude X-ray variability is accompanied by
substantially smaller changes at longer wavelengths. The UV
emission shows a possible long-term correspondence with the X-ray
behavior, being relatively faint during the X-ray low states and
increasing again during the recent X-ray recovery, whereas the
optical and infrared emission varies much more mildly. The
wavelength-dependent fractional variability also increases from
the infrared and optical bands toward the UV and X-ray regimes
(Figure~\ref{fig:Fvar}). The broadband SED further indicates that
the overall optical/UV emission from the accretion disk remains
comparatively stable during the X-ray faint state. These properties
suggest that the recurrent X-ray state transitions are primarily
associated with physical changes in the innermost disk--corona
system and/or ionized material along the line of sight, rather than
with a comparable global change in the accretion flow.

Previous X-ray studies have revealed strong reflection signatures
in Mrk~382. \citet{Tortosa2023} reported relativistic reflection
features, while the recent broadband \xmm--\textit{NuSTAR} analysis
of \citet{xu2025} showed that relativistic reflection can reproduce
the soft X-ray excess, broad Fe K$\alpha$ emission, and hard X-ray
Compton hump. In particular, \citet{xu2025} found substantial
differences in the reflection properties between the high- and
low-flux spectra and interpreted them, within a relativistic
reflection framework, as possible changes in the height of the hot
corona above the accretion disk.

Our revised spectral analysis broadly supports this picture while
also accounting for the additional absorption and distant-reflection
components indicated by the spectra. The disk ionization parameter
decreases from
$\log\xi_{\rm disk}=2.6^{+0.1}_{-0.3}$ in XM\_H to
$1.6\pm0.1$ in XM\_L, while the fitted reflection fraction increases
from $R_{\rm refl}=4.4^{+1.8}_{-1.1}$ to
$30.2^{+6.7}_{-6.4}$ (Table~\ref{tab:xspec_90pct}). The spectral
decomposition shown in Figure~\ref{fig:bestfit} illustrates this
change directly: in XM\_H, both the direct coronal continuum and
relativistic reflection contribute substantially, whereas in XM\_L
the direct continuum is strongly suppressed relative to the
relativistic reflection component. Importantly, the low-flux
spectrum remains strongly reflection dominated after both
photoionized absorption and distant reflection are included.
Thus, the principal spectral difference between XM\_H and XM\_L
cannot be removed simply by introducing these additional components.

Within a relativistic-reflection interpretation, this behavior is
consistent with changes in the geometry of the X-ray-emitting
corona. If the corona becomes more compact and/or decreases in
height above the disk, gravitational light bending can redirect a
larger fraction of the coronal radiation toward the inner disk
while reducing the fraction that reaches the distant observer
\citep[e.g.,][]{MiniuttiFabian2004}. The directly observed
continuum can consequently decline while the relative contribution
of reflection increases. Changes in coronal height have
specifically been proposed by \citet{xu2025} to explain the
different high- and low-flux states of Mrk~382. We emphasize,
however, that our use of \texttt{relxill} does not directly measure
the coronal height, and the large reflection fraction should
therefore be regarded as being consistent with, rather than uniquely
demonstrating, a more compact coronal configuration.

Changes in geometry alone may not provide a complete description of
the long-term variability. The possible coordinated UV and X-ray
evolution provides an additional constraint. Although the available
UV coverage is sparse and heterogeneous, the UV emission tends to
be fainter during the X-ray low states and becomes brighter again
during the recent X-ray recovery. We therefore regard the UV--X-ray
correspondence as qualitative rather than as a statistically
established correlation. Nevertheless, its direction is relevant:
in a simple light-bending picture with approximately constant
intrinsic coronal power, stronger light bending during the X-ray
faint state would redirect a larger fraction of the coronal
luminosity toward the disk, and therefore would not necessarily
produce a simultaneous decrease in the UV emission.

Recent studies have emphasized that UV/X-ray variability can reflect
a combination of disk reprocessing, intrinsic accretion-flow
fluctuations, and changes in the corona itself. Similar
behavior has been proposed in a number of highly variable AGNs
and NLS1 galaxies, where rapid changes in coronal geometry,
compactness, or energy dissipation can strongly modulate the
observed X-ray/UV variability
(e.g., \citealt{Ricci2020,Wu2020,Papoutsis2026}). A similar
possibility is attractive for Mrk~382. The large state-dependent
change in reflection fraction points to a change in how the coronal
radiation is distributed between the disk and the observer, while
the possible same-direction long-term UV and X-ray evolution may
indicate a simultaneous change in the intrinsic power dissipated
in the corona. A decrease in intrinsic coronal power during XM\_L,
together with stronger light bending, could reduce both the directly
observed X-ray continuum and the irradiating luminosity available
for disk reprocessing. Conversely, an increase in coronal power
during the recovery phase could contribute to the simultaneous
recovery of the X-ray and UV emission. Thus, both the geometry and
energetics of the corona may vary between the different X-ray
states of Mrk~382.

This interpretation remains tentative because the present UV
sampling does not permit a quantitative reverberation or lag
analysis. Moreover, the UV variability need not arise entirely from
X-ray reprocessing; intrinsic disk fluctuations and their inward
propagation may also contribute to the long-term UV/X-ray behavior
(e.g., \citealt{Hagen2024}). The current observations therefore
do not allow the relative contributions of intrinsic disk
variability, reprocessing, and changes in coronal energetics to be
separated quantitatively.

Ionized absorption provides an additional contribution to the
spectral variability. Strong and rapidly variable X-ray weakness
produced by ionized or partially covering material has been reported
in highly accreting AGNs, where clumpy disk winds can substantially
attenuate the X-ray continuum while producing much weaker changes
at optical/UV wavelengths
\citep[e.g.,][]{Huang2023,Huang2026}. Motivated by the soft-X-ray
residuals in Mrk~382, we explicitly tested this possibility using
XSTAR photoionization models. In XM\_L, the absorber is relatively
well constrained, with
$N_{\rm H,warm}\simeq1.9\times10^{21}\ {\rm cm^{-2}}$ and
$\log\xi_{\rm warm}\simeq1.6$. The absorber accounts for part of
the structured soft-X-ray residuals, demonstrating that ionized
absorption contributes to the detailed spectral shape. However,
the inferred column density is relatively modest and is unlikely
by itself to produce the factor-of-several broadband soft-X-ray
suppression between XM\_H and XM\_L.

The absorber parameters in XM\_H are considerably less well
constrained. In the final fit, the column density reaches the upper
boundary of the XSTAR grid, while the ionization parameter has a
large uncertainty. Tests with different initial parameter values
also reveal substantial degeneracy among the absorber parameters.
We therefore refrain from assigning a quantitative physical
interpretation to the high-state XSTAR solution or from using it to
infer a systematic change in absorber properties between XM\_H and
XM\_L. The robust conclusion from the absorption modeling is instead
that ionized material can contribute to the detailed soft-X-ray
spectral complexity, while the available spectra do not support
absorption as the sole origin of the large-amplitude state
transitions.

Ionized absorption may also affect the interpretation of the Fe-K
band, because highly ionized gas can introduce absorption structure
that modifies the apparent broad-band curvature
\citep[e.g.,][]{middei2023}. In XM\_L, however, the inclusion of
the XSTAR absorber does not remove the strong relativistic-reflection
component. The remaining absorption-like feature near the Fe-K band
is only marginally improved by an additional absorption line and is
therefore treated as tentative. At the same time, the narrow
Fe K$\alpha$ emission requires a contribution from distant,
weakly ionized reflecting material. We therefore include a separate
\texttt{xillver} component, whose ionization parameter is fixed at
$\log\xi=0$ after reaching the lower model boundary. The distant
reflection component is well constrained in XM\_L but is consistent
with zero within the uncertainties in XM\_H. Importantly, after both
warm absorption and distant reflection are included, XM\_L remains
strongly reflection dominated.

The multiwavelength behavior provides a complementary constraint.
Although the X-ray emission is substantially suppressed during the
low state, Mrk~382 remains within the normal range of the
$\alpha_{\rm ox}$--$L_{2500\text{\AA}}$ relation and does not enter
the canonical X-ray-weak regime. Recent studies suggest that
$\sim30\%$ of intermediate-redshift, highly accreting
($\lambda_{\rm Edd}\gtrsim1$) quasars may be intrinsically X-ray weak
by factors of $\sim10$ \citep[e.g.][]{Laurenti2022}. In many such systems, the X-ray weakness
has been interpreted as the result of attenuation by high-column-density,
often ionized gas along the line of sight, such as shielding material
associated with disk winds \citep[e.g.][]{Wu2012, Luo2015, Huang2023}. Many X-ray-weak
AGNs show dramatic X-ray variability with relatively little
corresponding UV variability, which differs from the behavior
observed in Mrk~382 (e.g., \cite{Huang2026, yangxh2026}). In Mrk~382, however, the
modest absorbing column inferred in XM\_L, together with the
subsequent recovery to an X-ray-bright state, argues against
persistent high-column-density obscuration as the dominant origin
of the recurrent faint states. Together with the correlated UV/X-ray
variability and the reflection-dominated low-state spectrum, these
results suggest that intrinsic changes in the compact corona are
more likely to dominate the observed variability behavior.

Taken together, the current observations favor a picture in which
the recurrent X-ray state transitions of Mrk~382 are primarily
associated with a dynamically evolving inner disk--corona system,
while ionized absorption contributes to the detailed spectral
complexity. The strong state-dependent change in reflection
fraction suggests changes in coronal geometry, consistent with the
recent reflection analysis of Mrk~382 by \citet{xu2025}. The
possible coordinated long-term UV and X-ray variability further
suggests that geometrical changes may be accompanied by changes in
the intrinsic coronal power, similar to the dynamically evolving
corona considered by \citet{Papoutsis2026}. We therefore suggest
that both the geometry and energetics of the corona may vary during
the recurrent X-ray state transitions. Variable ionized absorption
may modulate the observed spectrum, but the relatively modest
absorbing column measured in XM\_L does not favor absorption as the
sole driver of the large-amplitude flux changes.

\subsection{Implications for Changing-look and Frozen-look AGNs}

Mrk~382 provides an interesting comparison with both
classical CL AGNs and the so-called
``frozen-look'' AGNs. Classical CL-AGNs exhibit dramatic
continuum variability accompanied by the emergence or
disappearance of broad optical emission lines, commonly
interpreted as evidence for substantial changes in the
inner accretion flow or accretion state. Recent studies
have shown that such CL behavior can also occur in
high-accretion NLS1 galaxies, suggesting that NLS1s may
represent an important subclass of highly variable AGNs
operating close to the Eddington limit
\citep{xu2024}.

However, the physical connection between NLS1 galaxies and
classical CL-AGNs remains uncertain. Although CL events
have now been identified in several NLS1 galaxies,
recent studies suggest that the CL transition
may preferentially occur in systems with relatively low
Eddington ratios and accretion efficiencies
\citep{wangs2024}. In addition, the critical
accretion threshold required to fully suppress the
broad-line region may depend on black hole mass, with
lower-mass systems requiring comparatively higher
Eddington ratios \citep{Guo2025}. These results imply
that extreme variability in highly accreting NLS1
galaxies may not necessarily lead to classical optical
CL transitions, but instead could primarily
reflect structural changes in the compact corona and
innermost accretion flow.

In this context, the absence of clear optical
spectral-type changes in Mrk~382 provides an important
contrast to classical CL-AGNs. Although the source
exhibits repeated X-ray state transitions together with
significant UV variability, the optical continuum and
broad-line properties remain relatively stable based on
the currently available observations. This behavior
suggests that the observed variability is primarily
associated with changes in the compact corona or
innermost accretion flow, without triggering a global
accretion-state transition capable of producing a full
optical CL event.

In this respect, Mrk~382 appears more similar to the
``frozen-look'' AGNs, such as Mrk~335, in which extreme
X-ray variability occurs without corresponding dramatic
changes in the optical broad-line spectrum. Previous
studies of Mrk~335 have shown that the X-ray and
optical--UV emission are often only weakly correlated,
while the high-ionization broad He\,{\sc ii} emission line
does not respond strongly to simultaneous large-amplitude
X-ray variability
\citep{komossa2020,tripathi2020}. Such behavior has often
been interpreted as evidence for partial-covering,
dust-free absorption. However, other studies instead favor
intrinsic changes in the corona and inner accretion flow
\citep{gallo2018,Gallo2019}. Similar phenomenology may
also be present in Mrk~382.

At present, it remains unclear whether frozen-look AGNs
represent systems with intrinsically different variability
mechanisms, or whether some of them are simply CL-AGNs
observed with insufficient temporal cadence to capture
optical spectral transitions. In this context, Mrk~382
provides an important nearby laboratory for investigating
the physical connection between extreme X-ray variability,
ultraviolet/optical responses, and the possible emergence
of CL behavior in highly variable Seyfert
galaxies.

\section{Conclusions and Future Work}\label{sec:C}

We have presented a multi-epoch, multiwavelength study of the nearby NLS1 galaxy
Mrk~382 using observations from \textit{Swift}, \textit{Chandra},
\textit{XMM-Newton}, and \textit{eROSITA}, together with archival ultraviolet,
optical, and infrared data. We found that Mrk~382 exhibits recurrent extreme
X-ray variability, with repeated transitions between bright, intermediate,
and faint states over a timescale of $\sim15$ yr, while remaining in a
relatively bright X-ray state during most recent observations. The soft X-ray
flux varies by nearly an order of magnitude, whereas the ultraviolet emission
exhibits more moderate variability and a possible long-term
correspondence with the X-ray variations. The optical bands also show clear
long-term variability, although the variations are smoother and less pronounced
than those observed in the X-ray and ultraviolet bands. In contrast, the
mid-infrared variability remains relatively modest over the same period.

Detailed X-ray spectral modeling indicates that the low-flux state is
associated with a harder spectrum and enhanced reflection features, including
a prominent narrow Fe K$\alpha$ emission line. The low-flux spectrum
remains strongly reflection dominated after accounting for distant reflection
and ionized absorption, while the inferred warm absorber has a relatively
modest column density. These results suggest that ionized absorption may
contribute to the detailed spectral complexity but is unlikely to account for
the full amplitude of the X-ray state transitions alone. The pronounced change
in reflection strength between the high- and low-flux states instead supports
substantial changes in the inner disk--corona system. Together with the possible
coordinated long-term UV and X-ray variability, this suggests that both the
geometry and energetics of the corona may vary between different flux states.
The broadband SED further shows that the optical/UV accretion-disk continuum
remains largely intact even during the X-ray faint state. At the same time, the broadband SED indicates that, although the X-ray emission of
Mrk~382 is substantially suppressed in the low state, it does not reach the
canonical X-ray-weak regime.

Although Mrk~382 exhibits repeated dramatic X-ray state transitions, we do not
detect clear optical spectral-type changes based on the currently available
data. In this respect, Mrk~382 appears more similar to the so-called
``frozen-look'' AGNs, such as Mrk~335, than to classical CL AGNs. Recent
studies have suggested that CL phenomena may also occur in high-accretion NLS1
galaxies operating close to the Eddington limit. Mrk~382 therefore provides an
important nearby laboratory for investigating the connection between extreme
X-ray variability, ultraviolet/optical responses, and the possible emergence
of CL behavior in rapidly accreting Seyfert galaxies.

Future coordinated multiwavelength monitoring will be crucial for understanding
the physical origin of the recurrent X-ray transitions in Mrk~382. In
particular, dense and simultaneous X-ray and UV monitoring can
constrain the characteristic timescales and amplitudes of the state transitions
and test whether the apparent long-term UV--X-ray correspondence is
associated with disk reprocessing, intrinsic accretion-flow variations, or
changes in coronal energetics, while contemporaneous optical spectroscopy will
be essential for testing whether optical CL behavior eventually emerges. High-resolution X-ray spectroscopy during future low states will also
be valuable for constraining the ionization and velocity structure of the
absorbing gas, while X-ray reverberation measurements can provide an independent
probe of changes in coronal geometry. High signal-to-noise X-ray observations
during future low states will further help distinguish between coronal variability and variable-absorption scenarios through detailed
reflection and timing analyses. Such observations will provide important
constraints on the structure and evolution of the inner accretion flow in
highly variable NLS1 galaxies.

\section*{Acknowledgments}
We acknowledge the supports from National Natural Science Foundation of China (NSFC; grant Nos.12573110, 12133001), the Shenzhen Science and Technology Program (JCYJ20230807113910021) and the Natural Science Foundation of Top Talent of SZTU(GDRC202208).

Spectroscopic observations were obtained at Calar Alto Observatory
using the 2.2 m telescope equipped with the Calar Alto Faint Object
Spectrograph (CAFOS). We thank the Calar Alto Observatory staff for
their support during the observations. We acknowledge the support of
the staff of the Lijiang 2.4 m telescope. Funding for the telescope has
been provided by the Chinese Academy of Sciences (CAS) and the People's
Government of Yunnan Province.

This work made use of public data from PS1, NEOWISE, and ZTF. This
research also made use of observations from \textit{XMM-Newton},
\textit{Swift}, \textit{eROSITA}, and \textit{Chandra}. We acknowledge
the use of the HEASoft software package, provided by NASA's High Energy
Astrophysics Science Archive Research Center (HEASARC).

\textbf{\software{
SAS \citep{gabriel2004},
CIAO \citep{fruscione2006},
HEASoft \citep{2014ascl.soft08004N},
XSPEC \citep{Arnaud1996},
Astropy \citep{2013A&A...558A..33A,
2018AJ....156..123A,
2022ApJ...935..167A},
NumPy \citep{walt2011},
Matplotlib \citep{hunter2007}
}}

\clearpage   
\end{document}